\documentclass[a4paper]{jpconf}
\usepackage{graphicx}
\begin{document}

\title{Modeling the optical/UV polarization while flying around the tilted outflows of NGC~1068}
\author{Fr\'ed\'eric Marin$^1$, Ren\'e Goosmann$^1$ and Michal Dov{\v c}iak$^2$}
\address{$^1$Observatoire astronomique de Strasbourg, Section Hautes Energies, 11 Rue de l'Universit\'e,\\
F-67000 Strasbourg, France \\
$^2$ Astronomical Institute, Academy of Sciences of the Czech Republic, Bo{\v c}n\'i II 1401,\\
CZ-14131 Prague, Czech Republic}
\ead{frederic.marin@astro.unistra.fr}

\begin{abstract}
Recent modeling of multi-waveband spectroscopic and maser observations suggests that the ionized outflows in the nuclear region 
of the archetypal Seyfert-2 galaxy NGC 1068 are inclined with respect to the vertical axis of the obscuring torus. 
Based on this suggestion, we build a complex reprocessing model of NGC~1068 for the optical/UV band. 
We apply the radiative transfer code {\sc stokes} to compute polarization spectra and images. The effects 
of electron and dust scattering and the radiative coupling occurring in the inner regions of the multi-component object 
are taken into account and evaluated at different polar and azimuthal viewing angles. The observed type-1/type-2 polarization 
dichotomy of active galactic nuclei is reproduced. At the assumed observer's inclination toward NGC~1068, 
the polarization is dominated by scattering in the polar outflows and therefore it indicates their tilting angle with respect 
to the torus axis. While a detailed analysis of our model results is still in progress, we briefly discuss how they relate to 
existing polarization observations of NGC~1068.
\end{abstract}


\section{Introduction}

According to the unified model of active galactic nuclei (AGN) \cite{Antonucci1993} the accreting supermassive black hole of 
so-called type-2 thermal AGN lies hidden behind an optically thick, dusty torus. The obscuration prevents the broad line region 
to be seen and therefore type-2 AGN lack the kineticly broadened Balmer emission lines that characterize type-1 objects. 
Observations indicate that outflows are ejected from close to the accretion disk and it is possible that the torus funnel 
collimates these winds along polar directions. The launch of the wind is not yet understood, but it is a natural assumption 
that the ejection and collimation happen symmetrically with respect to the axis of the dusty torus and the accretion disk. 
In the case of the archetypical Seyfert-2 galaxy NGC~1068, the AGN is observed at a rather high inclination and the ionized 
winds are therefore well seen from the side. 

A recent study by \cite{Raban2009} suggests that the polar winds of NGC~1068 are inclined with respect to the torus axis. 
Based on this assumption, \cite{GoosmannMatt2011} and \cite{Goosmann2011} started to investigate the radiative transfer 
effects of the misaligned outflows on the induced X-ray polarization. The modeling suggests that the position angle of 
the wind on the sky plane could be determined unambiguously from a soft X-ray polarization measurement. 
In addition to this, the tilting angle of the wind with respect to the torus axis could be constrained by the rotation 
of the polarization position angle between the soft and hard X-ray bands.

The predictions for future X-ray polarization measurements are more solid when they can be coherently related to present-day polarization 
results from a different waveband. Therefore, we here present some results for the optical/UV counterpart of the same model setup. 
Similarly to \cite{Goosmann2011}, we present spectral and polarization modeling of NGC~1068 at several polar and azimuthal viewing angles. 
We use the new polarization mapping capacity that was recently implemented in our code {\sc stokes}. A more detailed description of the new 
{\sc stokes} version and more results of the current study will be published elsewhere (Marin et al., in preparation).


\section{The model setup for NGC~1068}

This research note focuses on the expected polarization signatures of NGC~1068 with misaligned outflows as a function 
of the polar and azimuthal viewing angle. We base our model on the description given in \cite{Raban2009} assuming that 
the bi-conical outflows are inclined by 18$^\circ$ from the vertical torus axis. The model setup is very similar to 
the one used in \cite{Goosmann2011}. Its main characteristics are summarized in Table \ref{Table1}. 
Note that here we did not include an irradiated accretion disk as it was done for the X-ray band where a 
lamp-post geometry was assumed. Instead, the negligibly small accretion disk produces an optical/UV continuum 
spectrum having a $F_*(E) \propto E^{-\alpha}$ power law shape. The spectral index is set to $\alpha = 1$. 
The lower and upper limits of the spectrum are set to 2000 \AA~and 8000 \AA, respectively. The Cartesian coordinate 
system of the model space and its relation to the polar and azimuthal viewing angle is as described in \cite{Goosmann2011}:
all reprocessing regions are centered on the coordinate origin and the wind axis is tilted inside the $yz$-plane 
leaning towards the positive $y$-axis. The observer's inclination and the tilting angle of the outflows are measured 
from the $z$-axis. The azimuthal viewing angle, $\phi$, is taken inside the $xy$-plane with respect to the 
negative $y$-axis. The modeling is carried out using the latest version of the Monte Carlo radiative 
transfer code {\sc stokes} presented in \cite{Goosmann2007} and recently upgraded (Marin et al., in preparation).

\begin{table}[]
 \begin{center}
{\footnotesize
   \begin{tabular}{|c|c|c|}
   \hline
      {\bf flared disk}                & {\bf dusty torus}                      & {\bf polar outflows}\\
   \hline
      $R_{\rm min} = 0.02$ pc          & $R_{\rm min} = 0.1$ pc                 & $R_{\rm min} = 0.3$ pc\\
      $R_{\rm max} = 0.04$ pc          & $R_{\rm max} = 0.5$ pc                 & $R_{\rm max} = 1.8$ pc\\
      half-opening angle = 20$^\circ$  & half-opening angle = 60$^\circ$        & half-opening angle = 40$^\circ$, tilted by 18$^\circ$\\
      equat. optical depth = 1         & equat. optical depth = 750             & vertical optical depth = 0.03\\
      electron scattering              & Mie scattering                         & electron scattering\\
   \hline
   \end{tabular}
}
  \caption{Parameters of the three scattering regions. The polar outflows and torus half-opening angles 
	   are measured with respect to the vertical symmetry axis of the torus. The half-opening angle 
	   of the flared disk is measured from the equatorial plane.}
  \label{Table1}
 \end{center}
\end{table}

We investigate the complex radiative coupling for a model composed of : (1) an optically and geometrically 
thick dusty torus, (2) an hourglass shaped, bi-conical polar wind filled with electrons, and (3) an equatorial, 
electron filled, radiation supported, flared disk. The equatorial scattering region is important in order to 
explain the observed polarization dichotomy between type-1 and type-2 AGN: it turns out from the observations 
that in type-1 AGN the polarization position angle in the optical/UV preferentially aligns with the 
projected symmetry axis of the polar outflows \cite{Antonucci1982}. Observationally, 
this axis is assumed to be traced by the small-scale radio jets (note that in radio-quiet AGN these jets are much 
smaller and weaker than the ballistic, kpc-size jets seen in radio-loud objects). The small jets in radio-quiet AGN 
should indicate nuclear outflows that contribute to the larger-scale, bi-conical ionization regions seen in optical/UV imaging. 
When investigating the scattering-induced polarization produced by the different reprocessing regions listed above, 
it turns out that it is difficult to reproduce the observed aligned polarization at type-1 viewing-angles, 
unless an equatorial scattering region, such as a flared disk, is present \cite{Chandra1960,Antonucci1984,Smith2002}.


\section{Spinning around NGC 1068 : modeling results}

The main goal in this research note is to revisit the X-ray polarimetry modeling of \cite{Goosmann2011} while 
shifting the energy band to UV and optical wavelengths and without doing any morphological changes to the model 
(with the exception of leaving aside the lamp-post irradiation geometry used only in the X-ray modeling).

\subsection{Spectropolarimetric signature}

\begin{figure}[htb]
  \centering
  \includegraphics[trim = 13mm 0mm 0mm 0mm, clip, width=47pc]{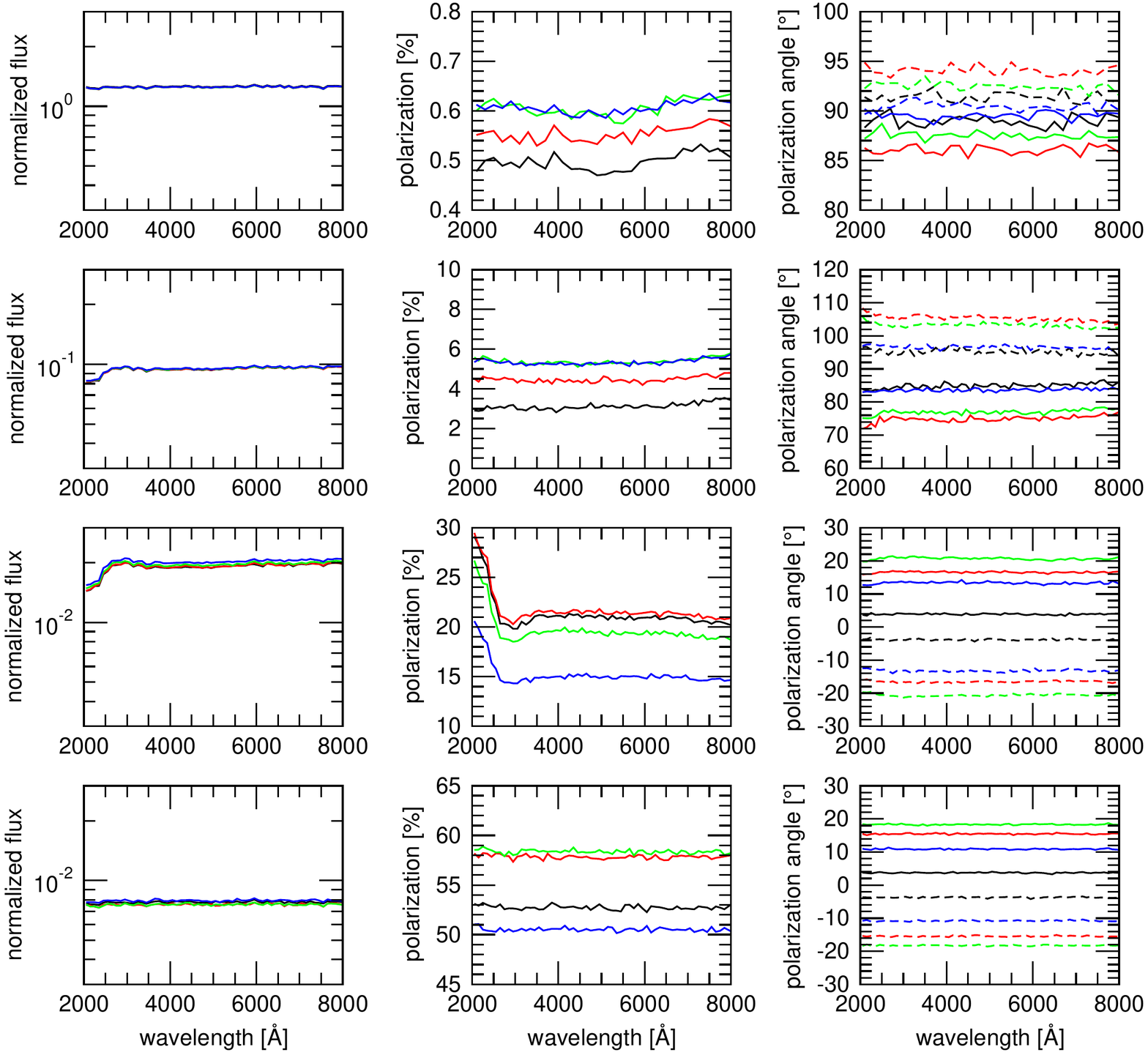}
  \caption{Spectral and polarization results in the optical/UV for the NGC~1068 model with inclined polar outflows. From top to 
	   bottom the panels represent the four inclinations $i \sim$ 26$^\circ$, 
	   $i \sim$ 60$^\circ$, $i \sim$ 73$^\circ$ and $i \sim$ 84$^\circ$. 
	   \textit{Left:} spectral flux $F/F_{cent}$ normalized to the source flux $F_{cent}$. 
	   \textit{Middle:} percentage of polarization $P$. \textit{Right:} polarization position angle $\psi$. 
	   The different curves denote four azimuthal viewing angles (half a round) measured from the 
	   negative $y$-axis : $\phi$ = 35$^\circ$ (blue), $\phi$ = 80$^\circ$ (green), 
	   $\phi$ = 125$^\circ$ (red) and $\phi$ = 170$^\circ$ (black). For the polarization position angle, $\psi$, 
	   the spectra at the symmetric azimuth ($\phi$ = -$\phi$) are also shown using dashed lines. 
	   For $\psi$ = 0$^\circ$ the polarization is perpendicular to the $z$-axis; it rotates clock-wise 
	   with rising $\psi$.}
  \label{Fig1}
\end{figure}

The spectropolarimetric modeling results are shown in Fig.\ref{Fig1}. We consider four inclinations at which we 
investigate the spectral and polarization response of the system. The angle $i \sim$ 26$^\circ$ features a type-1 object 
(face-on view), $i \sim$ 60$^\circ$ refers to a viewing angle passing through the upper layers of the torus, 
$i \sim$ 73$^\circ$ is around the expected inclination of NGC~1068 \cite{Honig2010} and, finally, $i \sim$ 84$^\circ$ 
features an extreme type-2 object (edge-on view). We cover half a round in azimuthal angle at $\phi$ = 35$^\circ$, 
$\phi$ = 80$^\circ$, $\phi$ = 125$^\circ$ and $\phi$ = 170$^\circ$. We extend the coverage to a full round for the 
spectra of the polarization position angle $\psi$. We define $\psi$ = 0$^\circ$ as ``perpendicular'' 
polarization with respect to the $z$-axis. It rotates clock-wise until $\psi$ = 90$^\circ$ (``parallel'' polarization).

Similarly to the X-ray results presented in \cite{Goosmann2011}, the spectral flux is rather $\phi$-independent at 
face-on view because the model does not deviate much from axis-symmetry. Since at low inclination the primary source 
is directly visible, the total flux is dominated by the power law continuum spectrum. Nevertheless, an additional scattered 
component contributes by more than ~20\% to the normalized flux. The flux level decreases rapidly as the observer's 
line of sight changes from type-1 to type-2 viewing directions. Heavy obscuration arises when the line of sight passes 
through the dusty torus, and the central source is no longer visible. Instead, scattered radiation dominates the spectrum.

The spectral shape of the polarization percentage, $P$, depends on the inclination: at $i \sim$ 26$^\circ$ and 
$i \sim$ 60$^\circ$, $P$ has low values due to strong dilution by the unpolarized primary radiation coming from the 
directly visible source. The low polarization is nearly wavelength-independent in the UV, which reveals electron scattering 
inside the flared disk as the main polarization mechanism. The polar outflows are seen either in transmission 
(forward-scattering in the near cone) or in straight-line reflection (backscattering in the far cone) and they thus do not 
contribute much to the polarization. A dim contribution of dust scattering to the polarization appears toward increasing 
wavelengths when the Mie scattering phase function becomes less anisotropic so that multiply scattered photons have a higher 
probability to escape from the torus funnel. When the line of sight crosses the torus horizon and moves towards the highest 
inclination, the effects of dust scattering are less prominent as the torus funnel becomes more and more obscured. Electron 
scattering in the polar outflows is dominant and therefore $P$ rises with $i$ and remains constant with wavelength. This is in 
agreement with the observations \cite{Miller1983, Antonucci1985, Antonucci1994}. The spectral shape of $P$ is nearly independent 
of azimuth, but not its normalization, which varies with $\phi$. 

The polarization position angle $\psi$ shows a systematic behavior as the virtual observer travels around the object 
in the $\phi$-direction. Despite the tilt of the outflows, the modeled $\psi$ confirms the observed polarization dichotomy at all $\phi$-angles: 
for type-1 objects, we obtain $\psi \sim 90^\circ$ and therefore the polarization roughly aligns with the axis of the outflow. For type-2 objects, the polarization is oriented perpendicularly to this axis. This picture would change, though, for more extreme tilting angles of the wind. The parallel polarization at type-1 viewing angles can only be preserved when the outflows are seen along a line of sight that mainly favors forward and backscattering. The polarization angle 
is symmetric in $\phi$, which is expected because the axis of the inclined outflows lies in the $yz$-plane. The variation of 
$\psi$ with $\phi$ is broader at type-2 than at type-1 view. By comparing the sign of $\psi$ at azimuthal viewing angles that are 
symmetric with respect to the $yz$-plane, we can recover the wind tilting direction: If the polarization angle is positive (negative), 
the polar outflows should be inclined in a clock-wise (counter clock-wise) direction from the $z$-axis.

The position of the outflow axis in NGC~1068 was investigated by \cite{Das2006} conducting kinematic modeling of the narrow 
line region. This region is assumed to be the natural continuation of the ionized winds. The kinematic modeling suggests that 
the axis does not lie in the plane of the sky but is indeed azimuthally rotated. From spectroscopy alone, however, 
this conclusion is difficult to verify and the polarization modeling carried out here can give an additional tracer of the 
scattering geometry. The combination of the polarization percentage and position angle adds two more observables that are 
sensitive to the position of the wind axis.

\subsection{Polarization imaging}

\begin{figure}[ht]
  \begin{minipage}{14pc}
    \includegraphics[trim = 8mm 5mm 0mm 10mm, clip, width=8cm]{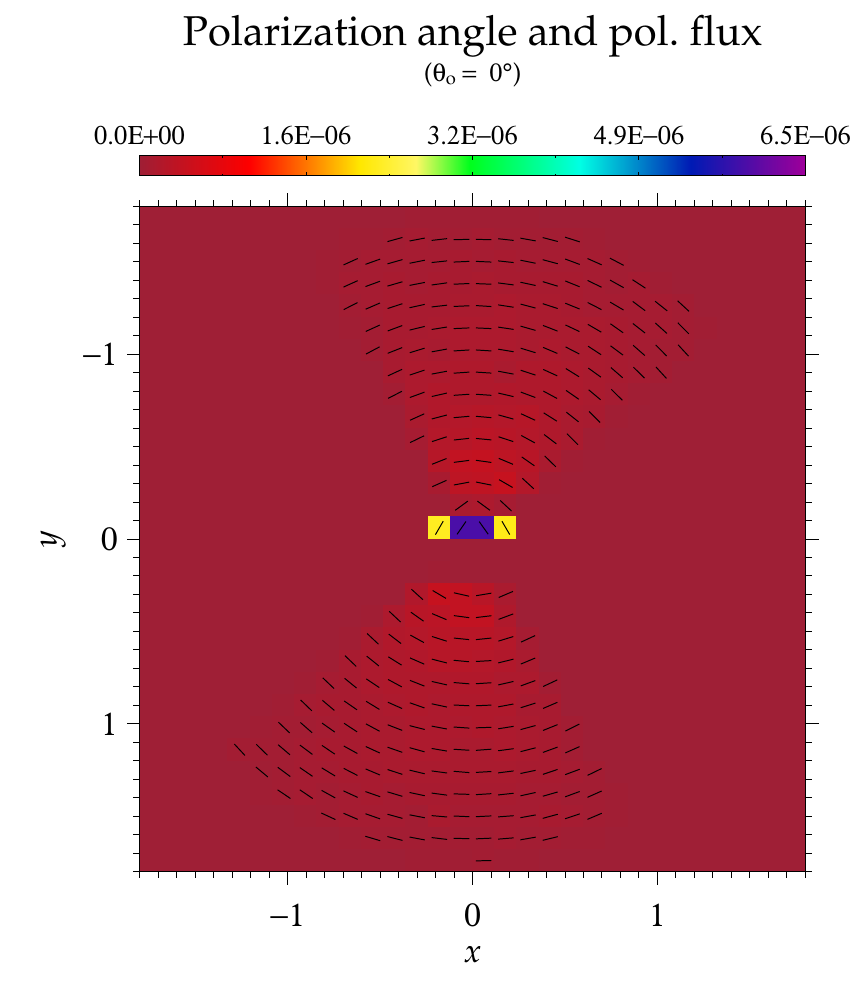}
    \includegraphics[trim = 8mm 5mm 0mm 10mm, clip, width=8cm]{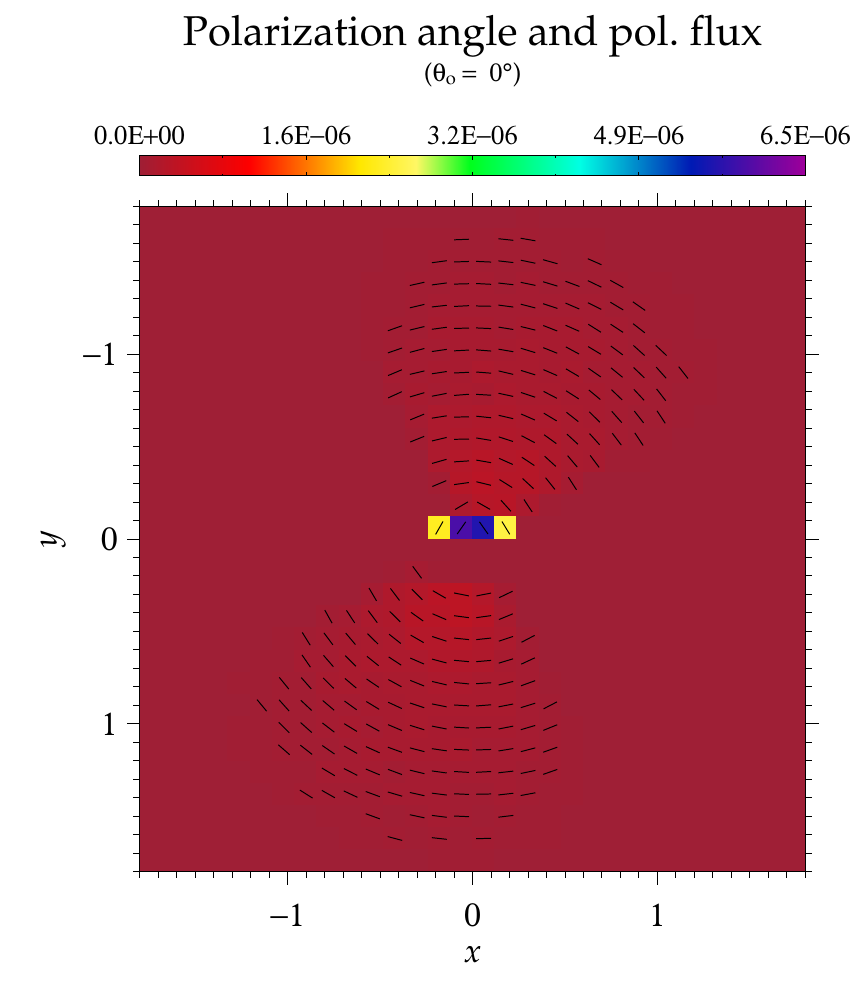}
  \end{minipage}\hspace{5pc}%
  \begin{minipage}{14pc}
    \includegraphics[trim = 8mm 5mm 0mm 10mm, clip, width=8cm]{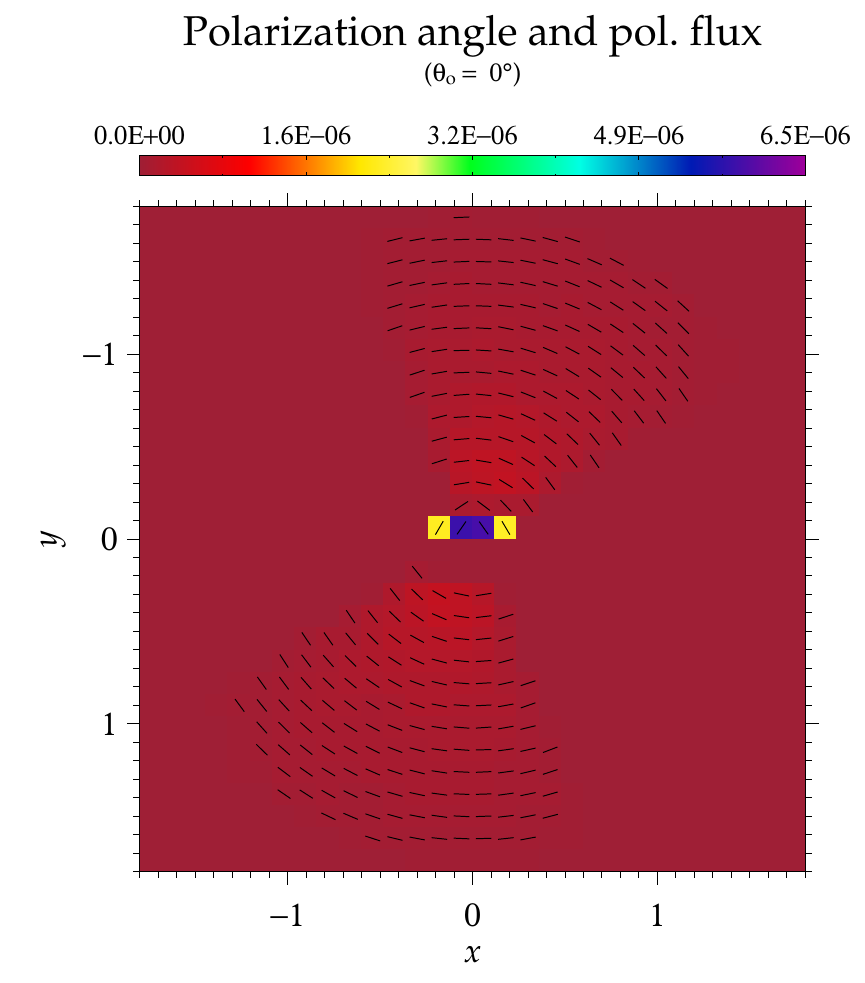}
    \includegraphics[trim = 8mm 5mm 0mm 10mm, clip, width=8cm]{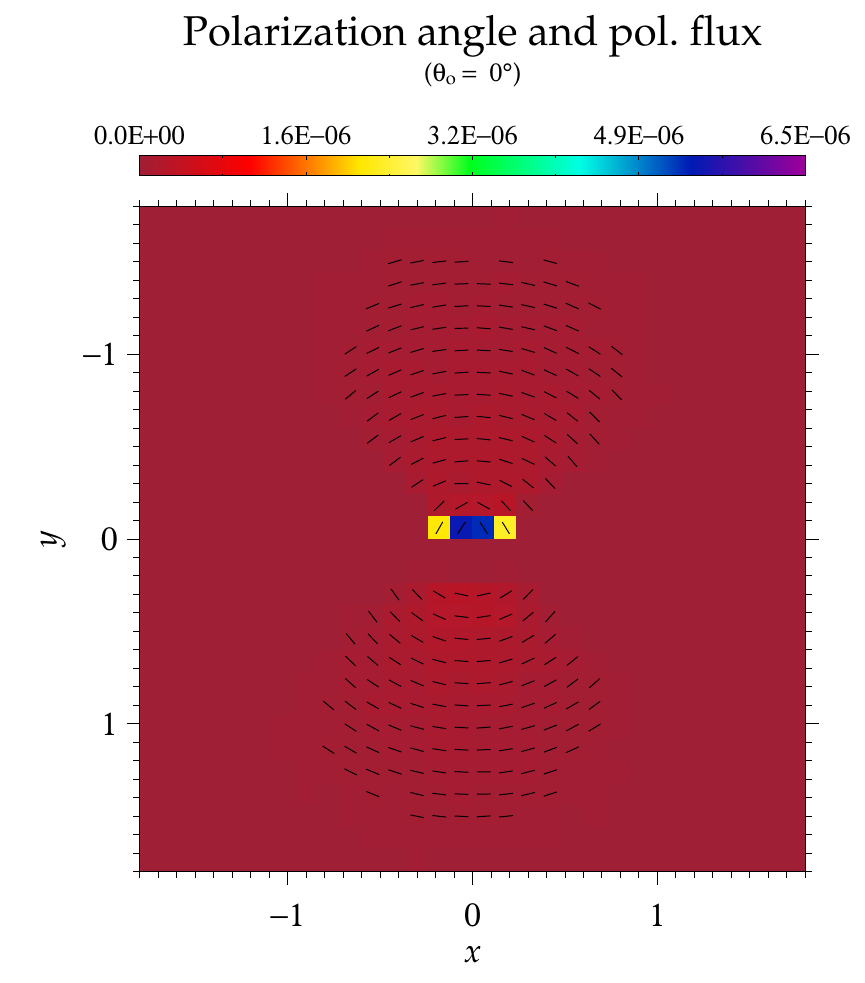}
  \end{minipage} 
   \caption{Modeling the spatial distribution of the polarized flux, $PF_{\nu}$, of our model for NGC~1068 with 
	    inclined polar outflows. The $PF_{\nu}$ is denoted by the color-coded grid and normalized to the total flux.
	    The polarization angle $\psi$ is represented by black bars (see text). The model region is observed at the 
	    inclination $i \sim$ 73$^\circ$ for the same $\phi$-angles as used in Fig.\ref{Fig1}. \textit{Top, left}: 
	    $\phi$ = 35$^\circ$, \textit{Top, right}: $\phi$ = 80$^\circ$, \textit{Bottom, left}: $\phi$ = 125$^\circ$, 
	    \textit{Bottom, right}: $\phi$ = 170$^\circ$.}
  \label{Fig2}
\end{figure}

Results of our simulated polarization cartography are given in Fig.~\ref{Fig2}. We show the maps at an inclination of $i \sim$ 73$^\circ$, 
which is in agreement with the expected viewing angle towards NGC~1068 as derived by \cite{Honig2010}. The maps simultaneously 
show the polarized flux, $PF_{\nu}$, and the polarization position angle, $\psi$. The angle $\psi$ is represented by black bars 
drawn in the center of each spatial bin. A vertical bar indicates a polarization of $\psi$ = 90$^\circ$, a bar leaning to 
the right denotes 90$^\circ > \psi > 0^\circ$ and a horizontal bar stands for $\psi$ = 0$^\circ$. For each pixel, the Stokes 
parameters are integrated over the full wavelength range.

The images illustrate how the net polarization of the AGN is determined by the polarized flux integrated over the projected extension 
of the object. Independently of the azimuthal angle, strong polarized flux with a polarization position angle geared towards a vertical 
direction is spotted at the center. The polarization here is dominated by the combined scattering in the flared electron disk and the 
funnel of the relatively flat torus. It was shown in \cite{Goosmann2007} that both such scattering regions produce parallel polarization. 
Even though the central polarized flux predominates over the average polarized flux from the outflows by a factor of 5, the large spatial 
extension of the latter determines the net polarization.

It is instructive to compare our model images with the observed UV polarization maps shown in Fig.~2 of \cite{Capetti1995}. Our model reproduces the arch-like distribution of polarization position angles across the extended outflows rather well. In the center, where according to \cite{Capetti1995} the main scattering source is located, the observed behavior still differs from our modeling. Although the observed polarization position angle near the central source shows some deviation from its behavior in the surroundings, this local turn is less prominent than our model suggests. It is possible that foreground matter on the observer's line of sight leads to additional polarization and can reduce the difference. Therefore, our analysis and modeling of the observed polarization images of NGC~1068 still needs some refinement, in particular in the central region of the object. It may also be that the inner surfaces of the torus are indeed steeper than assumed here so that the resulting polarization from the torus funnel is perpendicular and not parallel. Also, we have not investigated the impact of the electron optical depth in the polar outflows or the equatorial scattering disk yet. Such investigations are currently carried out and will be presented in a forthcoming paper (Marin et al., in preparation).

\section*{References}
\bibliography{iopart-num}

\end{document}